\documentclass[aps,prd,twocolumn,floatfix,tightenlines,groupedaddress,nofootinbib,amsmath]{revtex4}
\usepackage{epsfig}

\begin{document}

\title{$T$-violation in flavour oscillations as a test
for relativity principles at a neutrino factory}

\author{C. N. Leung}\email{leung@physics.udel.edu}\affiliation{
Department of Physics and Astronomy, University of Delaware,
Newark, Delaware 19716}

\author{Yvonne Y. Y. Wong}\email{ywong@physics.udel.edu}\affiliation{
Department of Physics and Astronomy, University of Delaware,
Newark, Delaware 19716}


\begin{abstract}
We study the effects of violation of the equivalence principle
(VEP) or violation of Lorentz invariance (LIV) in the neutrino
sector on the asymmetry between $T$-conjugate oscillation
probabilities, $\Delta P_T \equiv P(\nu_{\alpha} \to \nu_{\beta})
- P(\nu_{\beta} \to \nu_{\alpha})$, in a three-flavour framework.
We find that additional mixing due to these mechanisms, while
obeying all present bounds, can lead to an observable enhancement,
suppression, and/or sign change in $\Delta P_T$ for the preferred
energies and baselines of a neutrino factory.  The measurement of
this asymmetry can be used to establish a new upper limit of order
$10^{-26}$ on VEP or LIV in the $(\nu_e, \nu_\mu)$ and $(\nu_e,
\nu_\tau)$ sectors.
\end{abstract}

\maketitle

\section{Review and motivations}

The phenomenon of neutrino flavour oscillations follows from the
mixing of nondegenerate neutrino states.  In the ``standard''
scenario, a flavour neutrino is a linear combination of neutrino
mass eigenstates and the degeneracy is broken by means of small
neutrino masses. Alternative mechanisms that do not require
neutrino masses (or more precisely, nondegenerate neutrino masses)
abound.  One possibility is through a violation of the equivalence
principle (VEP) in the neutrino sector
\cite{bib:gasperini,bib:halprin}. Here, a flavour neutrino is a
mixture of neutrino gravitational eigenstates and
flavour-dependent neutrino-gravity couplings serve to lift the
degeneracy.

An essential phenomenological difference between the standard
and the VEP mechanisms is the manner in which the oscillation
probability $P(\nu_{\alpha} \to \nu_{\beta})$ depends on the
neutrino energy $E$. In the two-flavour case, a formal
transformation from the former to the latter scenario is readily
accomplished by replacing in $P(\nu_{\alpha} \to \nu_{\beta})$:
\begin{equation}
 \frac{\delta m^2}{2 E} \to 2 E |\phi| \delta \gamma,
\end{equation}
where $\delta m^2 \equiv m^2_2 - m^2_1$ is the neutrino squared
mass difference,  $\delta \gamma \equiv \gamma_2 - \gamma_1$ the
difference in the neutrino-gravity couplings, and $\phi$ the
Newtonian gravitational potential.  Note that a violation of
Lorentz invariance (LIV), where the breaking of the neutrino
degeneracy is assumed by dissimilar neutrino asymptotic speeds
$v_i$ \cite{bib:liv}, may also contribute to the oscillation
phenomenon with the same energy dependence as in the VEP case,
save for a minor notational change: $|\phi| \delta \gamma \to
\delta v/2$, where $\delta v \equiv v_2 - v_1$ \cite{bib:vepliv}.
Henceforth, we shall use only the VEP convention.

To date, the most comprehensive set of bounds on the VEP parameter
$|\phi| \delta \gamma$ comes from the CCFR experiment
\cite{bib:ccfr}: $|\phi \delta \gamma| \lesssim 10^{-22}$ for all
three  oscillation channels in the two-flavour limit
\cite{bib:pantaleone}. These extremely severe constraints have
since excluded VEP-LIV as an explanation to the LSND
$\overline{\nu}_{\mu}  \to \overline{\nu}_e$ transition puzzle
\cite{bib:lsnd} which requires $|\phi \delta \gamma| \gtrsim 5
\times 10^{-18}$ \cite{bib:hlp}, as well as
Mikheyev-Smirnov-Wolfenstein (MSW) type solutions (with $|\phi
\delta \gamma| \gtrsim 10^{-21}$
\cite{bib:hlp,bib:bahcall,bib:mansour}) to the solar neutrino
problem (see later).

In the case of atmospheric $\nu_{\mu}$ and $\overline{\nu}_{\mu}$
disappearance, the totality of the Super-Kamiokande atmospheric
neutrino data \cite{bib:superk} strongly favours an energy
dependence of $1/E^n$, where $n >0$, in the oscillation frequency
\cite{bib:pantaleone,bib:liparivep,bib:italian}. No acceptable
VEP-LIV solution exists for the complete set of contained,
partially contained, and upward going muon events. In an
interesting analysis, the authors of Ref.\ \cite{bib:italian}
considered VEP-LIV as a sub-leading effect to standard
mass-induced oscillations within a two-flavour $\nu_{\mu}
\leftrightarrow \nu_{\tau}$ framework, and derived the upper limit
$|\phi \delta \gamma| \lesssim 3 \times 10^{-24}$ at $90\ \%$ C.L.
for unconstrained VEP mixing in the $(\nu_{\mu}, \nu_{\tau})$
sector. For maximal VEP mixing, the bound on $|\phi \delta
\gamma|$ can be as stringent as $\sim 10^{-26}$.

On the other hand, it has been demonstrated in a number of recent
works \cite{bib:braziliansolar,bib:brazilian,bib:indian} that
vacuum oscillations induced by a ``just-so'' VEP with $|\phi
\delta \gamma| \sim 10^{-24}$ and $\sin^2 2 \theta \simeq 1$ can
still provide a good description of the solar neutrino data
\cite{bib:homestake,bib:sage,bib:gallex,bib:gno,bib:kamiokande,
bib:superksolar,bib:sno}.  The analyses of Ref.\
\cite{bib:brazilian} put the solution's goodness of fit at $60 \%$
C.L., a figure that is comparable even to the confidence level of
$75\ \%$ for the  standard large mixing angle (LMA) solution.
However, the LMA oscillation parameters have now been confirmed by
the terrestrial experiment KamLAND, through the observation of
reactor $\overline{\nu}_e$ disappearance over a typical baseline
of $180 \ {\rm km}$ \cite{bib:kamland}. While these new results do
not exclude VEP at the aforementioned scale, a pure VEP-LIV
solution to the solar $\nu_e$ deficit is  no longer justified.

It is clear from the above discussion that neutrino oscillations
provide very stringent tests for the principle of equivalence and
Lorentz invariance.  We examine in the present work the
possibility of further constraining VEP-LIV in the neutrino sector
with future experiments.  Currently operating long baseline
terrestrial experiments (e.g., K2K), those under construction
(e.g., MINOS), and prospective neutrino factories
\cite{bib:nufact} will provide for us a means to obtain more
precise information on the neutrino mixing properties, as well as
an opportunity to establish $CP$- or $T$-violation in the lepton
sector.  In particular, the capacity of a neutrino factory to
produce {\it both} electron and muon neutrino and antineutrino
beams in high intensities means that it is in principle possible
to directly compare the $T$-conjugate oscillation probabilities
$P(\nu_{\mu} \to \nu_e)$ and $P(\nu_e \to \nu_{\mu})$
\cite{bib:bueno}. Despite the associated experimental challenges,
this is theoretically the cleanest way to resolving any $CP$- or
$T$- violating phases in the neutrino mixing matrix: a nonzero
measurement of $\Delta P_T \equiv P(\nu_{\alpha} \to \nu_{\beta})
- P(\nu_{\beta} \to \nu_{\alpha})$ necessarily implies the
existence of such a phase, in contrast with the $CP$-asymmetry
$\Delta P_{CP} \equiv P(\nu_{\alpha} \to \nu_{\beta}) -
P(\overline{\nu}_{\alpha} \to \overline{\nu}_{\beta})$ which
inevitably contains additional contribution from terrestrial
matter effects.\footnote{It should be noted that $T$-violating
effects can arise from an earth matter distribution that is not
symmetric about the midpoint of the neutrino's trajectory,
independently of the intrinsic $CP$- or $T$-violating phase
\cite{bib:tviolation}.  See Ref.\ \cite{bib:lindner} for a recent
study.} Finally, a neutrino factory is ideal for this study since
it involves high energy neutrinos travelling a very long distance,
which tends to enhance the VEP-LIV effects \cite{bib:datta}.

Assuming three neutrino flavours and that standard oscillations
account for both the atmospheric neutrino anomaly and solar
neutrino deficit, we consider VEP-LIV as a sub-leading mechanism
(while respecting all present bounds), and examine its consequence
on the $T$-asymmetry $\Delta P_T$ for the design parameters of a
neutrino factory. We find that although the presently allowed
$|\phi| \delta \gamma$'s are far too weak to produce any  effect
on the actual oscillation probabilities beyond the few percent
level, the $T$-asymmetry can be significantly enhanced,
suppressed, or flip sign, depending on the VEP-LIV breaking scale
as well as the energy of the neutrinos.

We shall begin by introducing the parameters relevant for the
present phenomenological study

\section{Definitions of Phenomenological Parameters}
\label{sec:nomenclature}

Consider the evolution of a three-flavour system in a medium,
\begin{equation}
\label{eq:evolution}
 i \frac{d}{dt} \nu = \widetilde{H} \nu,
\end{equation}
where $\nu = (\nu_e, \ \nu_{\mu},\  \nu_{\tau} )^T$.   Suppose
that, in addition to the usual mass-mixing and matter potential
terms, $U_M M U_M ^{\dagger}$ and $V$, the total Hamiltonian
$\widetilde{H}$ governing this evolution contains also a third
term $U_G G U^{\dagger}_G$ arising from a violation of the
equivalence principle (VEP):
\begin{equation}
\label{eq:totalham}
 \widetilde{H}  = \frac{1}{2E} U_M M
 U_M^{\dagger} + 2 E \ U_G G U_G^{\dagger} + V,
\end{equation}
where
\begin{eqnarray}
 M &=& {\rm Diag} (m_1^2,\ m_2^2,\ m_3^2),\nonumber\\ G & =&
 |\phi| \times {\rm Diag}(\gamma_1, \ \gamma_2, \ \gamma_3),
 \nonumber \\ V& =& \sqrt{2} \ G_F N_e \times {\rm Diag}(1,\ 0, \
 0). \label{eq:mgamma}
\end{eqnarray}
Here, $m^2_{1,2,3}$ are the squared masses of the neutrino mass
eigenstates, $\phi$ is the gravitational potential,
$\gamma_{1,2,3}$ are the neutrino-gravity couplings, $E$ is the
neutrino energy, $G_F$ is the Fermi constant, $N_e$ is the
electron number density of the medium, and $U_{M,G}$ are two
independent unitary matrices relating the flavour eigenstates to,
respectively, the mass and gravitational eigenstates.  Note that
the subscript capital $M$ stands for ``mass'', and is not to be
confused with the commonly used small $m$ which denotes
``matter''.

A generic $3 \times 3$ unimodular unitary matrix may be
parameterised with three mixing angles and six phases:
\begin{equation}
 U \equiv e^{i \eta} \ {\rm Diag}(1,\ e^{i\alpha},\ e^{i \beta}) \
 U' \ {\rm Diag}(1, \ e^{i \xi}, \ e^{i \chi}),
\end{equation}
where $\eta$, $\alpha$, $\beta$,  $\xi$, and $\chi$ are five
arbitrary phases, and the matrix $U'$ contains the three mixing
angles and the sixth phase, i.e., the $CP$- or $T$-violating
phase, such as the familiar CKM matrix.\footnote{A common practice
is to express some of these extra phases in terms of $SU(3)$
generators, utilising the diagonal Gell-Mann matrices $\lambda_3$
and $\lambda_8$.  We find our simple parameterisation to be more
convenient for the present phenomenological study.} The phase
$\eta$ is an unobservable overall phase.  $\xi$ and $\chi$ may
correspond to physical Majorana phases (if neutrinos are Majorana
particles), or to unphysical phases (if neutrinos are Dirac
particles), neither of which contribute to flavour oscillations.
If only the mass term is present, the remaining two phases
$\alpha$ and $\beta$ are unphysical, and can always be absorbed
into the definitions of the flavour eigenstates, e.g.,
$\nu_{\mu}'= e^{-i \alpha_M} \nu_{\mu}$, leaving $U_M'$ the only
observable entity. Similarly, if $U_G G U^{\dagger}_G$ is the
unique term, only $U_G'$ contributes to the oscillation
phenomenon.  In the general mass plus gravity case, however, it is
not possible to completely eliminate the $\alpha$ and $\beta$
phases.  If we choose $U_M = U_M'$, it follows that the total
Hamiltonian $\widetilde{H}$ must take the form
\begin{eqnarray}
 \widetilde{H}  &=& \frac{1}{2E} U_M' M U_M'^{\dagger} + 2 E \ W
 U_G' G U_G'^{\dagger} W^{\dagger} + V \nonumber \\ &\equiv & H^M +
 W H^G W^{\dagger} + V \nonumber \\ &\equiv& H +  W H^G
 W^{\dagger},
\end{eqnarray}
where $W = {\rm Diag}(1,\ e^{i\alpha},\ e^{i \beta})$, with
$\alpha \equiv \alpha_G - \alpha_M$ and $\beta \equiv \beta_G -
\beta_M$, and we have defined $H \equiv H^M + V$ as the
``standard'' mass-induced mixing plus matter potential
sub-Hamiltonian for future reference. To transform from $H^G$ to
$W H^G W^{\dagger}$, one simply makes the following replacements:
\begin{eqnarray}
 H^G_{e \mu } \to   H^G_{ e \mu} e^{i \alpha},  \qquad & \quad  &
 H^G_{\mu e} \to   H^G_{\mu e} e^{- i \alpha}, \nonumber \\
 H^G_{e \tau} \to H^G_{e \tau} e^{i \beta}, \qquad  & \quad  &
 H^G_{\tau e} \to H^G_{\tau e}e^{- i \beta}, \nonumber \\
 H^G_{\mu \tau}  \to H^G_{\mu \tau} e^{i (\beta -\alpha)}, & \quad
 & H^G_{\tau \mu} \to H^G_{\tau \mu} e^{i (\alpha -\beta)}.
\end{eqnarray}
Thus the inclusion of a gravitational term in the total
Hamiltonian leads to, in principle, eight, not  six, extra
observable parameters!

In the following, we shall make the simplifying assumption that
the splitting between two of the three gravitational eigenstates
is too small to be probed by present day experiments and
experiments in the near future.  This near degeneracy means that
only one $|\phi| \delta \gamma$ and two VEP mixing angles
contribute to the observable part of $H^G$, and the $CP$- or
$T$-violating phase inherent in $U_G'$ can be absorbed into the
definitions of the phases $\alpha$ and $\beta$.  We shall adopt
the convention $\gamma_1 \simeq \gamma_2 $ and $\delta \gamma
\equiv \gamma_3 - \gamma_2$, and parameterise the effective $U_G'$
as
\begin{equation}
 U_G' = \left(\begin{array}{ccc}
            \cos \varphi & 0 & \sin \varphi \\
            -\sin \psi \sin \varphi & \cos \psi & \sin \psi \cos
            \varphi \\
            -\cos \psi \sin \varphi & - \sin \psi & \cos \psi \cos
            \varphi \end{array} \right),
\end{equation}
where $\varphi$ and $\psi$ run from $-\pi/2$ to $\pi/2$.  The
resulting three-flavour sub-Hamiltonian $H^G$ takes the form
\begin{equation}
\label{eq:threeflav}
 E |\phi| \delta \gamma \left(\begin{array}{ccc}
                    2 \sin^2 \! \varphi &  \sin \psi \sin 2
                    \varphi &  \cos \psi \sin 2
                    \varphi \\
                    \sin \psi \sin 2 \varphi & 2 \cos^2 \! \varphi
                    \sin^2 \! \psi & \cos^2 \!  \varphi  \sin 2 \psi \\
                    \cos \psi \sin 2 \varphi & \cos^2 \! \varphi \sin 2
                    \psi & 2 \cos^2 \!  \varphi  \cos^2 \!  \psi
                    \end{array}  \right),
\end{equation}
which reduces to a pure two-flavour description in terms of one
mixing angle when either $\varphi$ or $\psi$ is fixed at $0, \
\pi/2$, or $- \pi/2$.  Specifically
\begin{eqnarray}
 \sin^2 \! \psi=1 & \Rightarrow & \nu_e \leftrightarrow \nu_{\mu} \
 {\rm VEP \ mixing} , \nonumber \\ \sin^2 \! \psi=0 & \Rightarrow&
 \nu_e \leftrightarrow \nu_{\tau}\ {\rm VEP \ mixing},\nonumber \\
 \sin^2 \! \varphi = 0 &\Rightarrow& \nu_{\mu} \leftrightarrow
 \nu_{\tau} \ {\rm VEP \ mixing}.
\end{eqnarray}
Observe that the sign of the parameter $|\phi| \delta \gamma$ may
be positive or negative.  As we shall see, this has important
consequences when considered together with $H^M +V$.

For the mass-mixing term $H^M$, we adopt the usual MNS
parameterisation \cite{bib:MNS} for $U_M'$:
\begin{equation}
 U_M'\! = \! \left( \! \begin{array}{ccc}
                c_{1} c_{3} & s_{1} c_{3} & s_{3} e^{-i \delta} \\
                -s_{1} c_{2} - c_{1} s_{2} s_{3} e^{i \delta}
                    & c_{1} c_{2} - s_{1} s_{2} s_{3} e^{i \delta}
                    & s_{2} c_{3} \\
                s_{1} s_{2} - c_{1} c_{2} s_{3} e^{i \delta}
                    & -c_{1} s_{2} - s_{1} c_{2} s_{3} e^{i \delta}
                    & c_{2} c_{3} \end{array} \! \right),
\end{equation}
where $s_i \equiv \sin \theta_i$, $c_i \equiv \cos \theta_i$, $i=
1,2,3$, and $\delta$ is the $CP$- or $T$-violating phase.   The
solar and atmospheric oscillation parameters are identified as
\begin{eqnarray}
 \delta m^2_{\rm atm} = \delta m^2_{32} \equiv m^2_3 - m^2_2, &
 \quad & \theta_{\rm atm} \equiv  \theta_2, \nonumber \\ \delta
 m^2_{\rm sun} = \delta m^2_{21} \equiv m^2_2 - m^2_1, &\quad &
 \theta_{\rm sun} \equiv \ \theta_1,
\end{eqnarray}
assuming that maximal $\nu_{\mu} \leftrightarrow \nu_{\tau}$
oscillations with $\delta m^2_{\rm atm} \simeq 3 \times 10^{-3} \
{\rm eV}^2$, and the large mixing angle (LMA) solution ($\delta
m^2_{\rm sun} \simeq 5 \times 10^{-5} \ {\rm eV}^2$, $\sin 2
\theta_{\rm sun} \simeq 0.8$) explain, respectively, the
atmospheric and solar neutrino anomalies.  Given these solutions,
a three-flavour analysis of the CHOOZ data sees the remaining
angle $\theta_3$ constrained to $\tan^2 \theta_3 < 0.055$ at 90\%
C.L. \cite{bib:bestfits}.

\section{Why $T$-violation experiments?}

Solution to the Hamiltonian  (\ref{eq:totalham}) may be obtained
by first transforming to a new basis defined as
\begin{equation}
 \widetilde{U}^{\dagger} \widetilde{H} \widetilde{U} = {\rm
 Diag}(\widetilde{\lambda}_1, \ \widetilde{\lambda}_2, \
 \widetilde{\lambda}_3),
\end{equation}
where $\widetilde{\lambda}_{1,2,3}$ are the eigenvalues, and the
energy dependent unitary matrix $\widetilde{U}$ contains by
definition one independent effective $CP$- or $T$-violating phase.
Assuming constant $N_e$ and $\phi$, the asymmetry between the
$T$-conjugate oscillation probabilities
$\widetilde{P}(\nu_{\alpha} \to \nu_{\beta})$ and
$\widetilde{P}(\nu_{\beta} \to \nu_{\alpha})$ is given by
\begin{eqnarray}
\label{eq:deltapt}
 \Delta \widetilde{P}_{T} & \equiv &
 \widetilde{P}(\nu_{\alpha} \to \nu_{\beta}) -
 \widetilde{P}(\nu_{\beta} \to \nu_{\alpha}) \nonumber \\ &=&
 16 \widetilde{J} \ \sin \frac{\widetilde{\Delta}_{12} L}{2}\ \sin
 \frac{\widetilde{\Delta}_{23} L}{2}\ \sin
 \frac{\widetilde{\Delta}_{31} L}{2},
\end{eqnarray}
where $\widetilde{\Delta}_{ij} \equiv \widetilde{\lambda}_i -
\widetilde{\lambda}_j$, with $i,j=1,2,3$, is the difference
between the $i$th and the $j$th eigenvalues, and
\begin{equation}
\label{eq:jarlskog}
 \widetilde{J} \equiv \widetilde{J}_{\alpha
 \beta}^{ij} = {\rm Im}(\widetilde{U}_{\beta i}
 \widetilde{U}_{\beta j}^* \widetilde{U}_{\alpha i}^*
 \widetilde{U}_{\alpha j})
\end{equation}
is the effective Jarlskog factor \cite{bib:jarlskog} encapsulating
the said effective phase.

Explicit evaluation of  the quantities $\widetilde{J}$ and
$\widetilde{\Delta}_{ij}$ is a tedious, though straightforward,
task.  Fortunately, there exists a very useful identity which
relates them to the original Hamiltonian in flavour
space \cite{bib:harrison}:
\begin{equation}
\label{eq:harrisonscott}
 \widetilde{\Delta}_{12}
 \widetilde{\Delta}_{23} \widetilde{\Delta}_{31} \widetilde{J} =
 {\rm Im}(\widetilde{H}_{e \mu} \widetilde{H}_{\mu \tau }
 \widetilde{H}_{\tau e}).
\end{equation}
Observe that (i) the right hand side of Eq.\
(\ref{eq:harrisonscott}) is dependent only on the off-diagonal
elements of $\widetilde{H}$, and (ii) the left hand side of Eq.\
(\ref{eq:harrisonscott}) bears a striking resemblance to the right
hand side of Eq.\ (\ref{eq:deltapt}).   These immediately suggest
that the observable $\Delta \widetilde{P}_T $ is potentially
very sensitive to the additional mixing terms arising from VEP.

We do not consider the $CP$ asymmetry $\Delta \widetilde{P}_{CP}
\equiv \widetilde{P}(\nu_{\alpha} \to \nu_{\beta}) -
\widetilde{P}(\overline{\nu}_{\alpha} \to \overline{\nu}_{\beta})$
in the present work.  This measurable inevitably contains
additional contribution from terrestrial matter effects, thus
rendering its analytical description much more complicated.

\section{Analysis}

We now analyse the effects of VEP on the observed $T$-asymmetry.
Consider the ratio
\begin{equation}
\label{eq:ratio}
 \frac{\Delta \widetilde{P}_T}{\Delta P_T} =
 \frac{\widetilde{J} \ \sin
 \frac{\widetilde{\Delta}_{12} L}{2}\ \sin
 \frac{\widetilde{\Delta}_{23} L}{2}\ \sin
 \frac{\widetilde{\Delta}_{31} L}{2}}{J \ \sin \frac{\Delta_{12}
 L}{2}\ \sin \frac{\Delta_{23} L}{2}\ \sin \frac{\Delta_{31}
 L}{2}},
\end{equation}
where the quantities with a tilde are as defined in Eqs.\
(\ref{eq:deltapt}) and (\ref{eq:jarlskog}), and those without
correspond to their ``standard'' $H^G=0$ counterparts.  Using the
matter-invariants
\begin{equation}
 {\rm Im}(H_{e \mu} H_{\mu \tau } H_{\tau e}) =  {\rm Im}(H^M_{e
 \mu} H^M_{\mu \tau } H^M_{\tau e}),
\end{equation}
and
\begin{equation}
 J =  \frac{1}{8 E^3} \frac{\delta m^2_{12} \delta m^2_{23} \delta
 m^2_{31}}{\Delta_{12} \Delta_{23} \Delta_{31}} J^M,
\end{equation}
where
\begin{equation}
 J^M = s_1 c_1 s_2 c_2 s_3 c_3^2 \sin \delta
\end{equation}
is the Jarlskog factor in vacuum, we rewrite Eq.\ (\ref{eq:ratio})
as
\begin{eqnarray}
 \frac{\Delta \widetilde{P}_T}{\Delta P_T}  & = &  \frac{8 E^3
 \Delta_{12} \Delta_{23} \Delta_{31} \widetilde{J}}{\delta m^2_{12}
 \delta m^2_{23} \delta m^2_{31} J^M } \nonumber \\ && \times
 \left( \frac{\sin \frac{\widetilde{\Delta}_{12} L}{2}\ \sin
 \frac{\widetilde{\Delta}_{23} L}{2}\ \sin
 \frac{\widetilde{\Delta}_{31} L}{2}}{\sin \frac{\Delta_{12} L}{2}\
 \sin \frac{\Delta_{23} L}{2}\ \sin \frac{\Delta_{31} L}{2}}
 \right).
\end{eqnarray}
Rough estimates of the differences between the eigenvalues
$\widetilde{\Delta}_{ij},\ ij = 12, 23, 31$ may be extracted from
the Hamiltonian (\ref{eq:totalham}) and Eq.\ (\ref{eq:mgamma}):
\begin{eqnarray}
 \widetilde{\Delta}_{12}\! \! & \simeq & \!\!
 \frac{m^2_1\!-\! m^2_2}{2 E} + {\cal O}( \sqrt{2} G_F N_e) + {\cal
 O} (2 E |\phi| \delta \gamma),\label{eq:masses21} \\
 \widetilde{\Delta}_{23} \!\! & \simeq & \!\! \frac{m^2_2 \! -\!
 m^2_3}{2 E} + {\cal O} (2 E |\phi| \delta \gamma), \\
 \widetilde{\Delta}_{31} \!\! & \simeq & \!\!  \frac{ m^2_3 \! - \!
 m^2_1}{2 E} - {\cal O}( \sqrt{2} G_F N_e) + {\cal O} (2 E |\phi|
 \delta \gamma).\label{eq:masses}
\end{eqnarray}
Their standard counterparts $\Delta_{ij}$ are given similarly,
save for the absence of the ${\cal O}(2 E |\phi| \delta \gamma)$
terms.

Suppose now that
\begin{equation}
\label{eq:gammadelta}
 2 E |\phi \delta \gamma| \ll \frac{|\delta m^2_{32}|}{2E},
\end{equation}
or, equivalently,
\begin{equation}
 |\phi \delta \gamma| \ll 2.5 \times 10^{-19} \frac{|\delta
 m^2_{32}|}{{\rm eV}^2} \left( \frac{\rm GeV}{E} \right)^2.
\end{equation}
This immediately allows us to set $ \widetilde{\Delta}_{23} \simeq
\Delta_{23}$, and thus
\begin{equation}
\label{eq:delta32}
 \sin \frac{\widetilde{\Delta}_{23} L}{2} \simeq
 \sin \frac{\Delta_{23} L}{2}.
\end{equation}
For $|\delta m^2_{32}| \simeq 3 \times 10^{-3}\ {\rm eV}^2$, and a
neutrino factory producing a neutrino beam of average energy
$13 \ {\rm GeV}$, Eq.\ (\ref{eq:gammadelta}) is a good
approximation for $|\phi \delta \gamma| \ll 4 \times 10^{-24}$,
a region of parameter space well below the present CCFR bounds.
Further approximations are in line, depending on whether
$\widetilde{\Delta}_{31}$ is {\it on-} or {\it off-resonance}.

{\it Off-resonance}: $\widetilde{\Delta}_{31} \neq 0$.  If there
is no exact or near cancellation between  ${\cal O}( \sqrt{2} G_F
N_e)$  and other terms in Eq.\ (\ref{eq:masses}), the condition
(\ref{eq:gammadelta}) implies also that $ \widetilde{\Delta}_{31}
\simeq \Delta _{31}$, and therefore
\begin{equation}
 \sin \frac{\widetilde{\Delta}_{31} L}{2} \simeq
 \sin \frac{\Delta_{31}L}{2}.
\end{equation}
Hence,
\begin{equation}
\label{eq:deltapapprox}
 \frac{\Delta \widetilde{P}_T}{\Delta P_T}
 \simeq \frac{8 E^3 \Delta_{12} \widetilde{\Delta}_{23}
 \widetilde{\Delta}_{31} \widetilde{J}}{\delta m^2_{12} \delta
 m^2_{23} \delta m^2_{31} J^M } \frac{\sin
 \frac{\widetilde{\Delta}_{12} L}{2}}{\sin \frac{\Delta_{12} L}{2}},
\end{equation}
and only $\widetilde{\Delta}_{12}$ is substantially affected by
$H^G$.  Equation (\ref{eq:deltapapprox}) may be further simplified
if the neutrino energy and the experimental baseline are such that
\begin{equation}
 \sin \frac{\widetilde{\Delta}_{12} L}{2} \simeq
 \frac{\widetilde{\Delta}_{12} L}{2}, \qquad \sin
 \frac{\Delta_{12} L}{2} \simeq \frac{\Delta_{12} L}{2},
\end{equation}
or, roughly speaking, if the experimental baseline satisfies the
requirement
\begin{equation}
\label{eq:baselineoffres}
 L < \frac{2}{|\widetilde{\Delta}_{12}|} \simeq \frac{2}{{\rm
 Max}\left( \frac{\delta m^2_{21}}{2E},  \ \sqrt{2} G_F N_e, \ 2E
 |\phi \delta \gamma| \right)}.
\end{equation}
The outcome of this approximation is a $T$-asymmetry ratio that is
related in a simple way to the off-diagonal elements of the
original Hamiltonians in flavour space,
\begin{equation}
\label{eq:ratioapprox}
 \frac{\Delta \widetilde{P}_T}{\Delta P_T}
 \simeq \frac{8 E^3 \widetilde{\Delta}_{12} \widetilde{\Delta}_{23}
 \widetilde{\Delta}_{31}\widetilde{J}  }{\delta m^2_{12} \delta
 m^2_{23} \delta m^2_{31} J^M}
 =
 \frac{{\rm Im}(\widetilde{H}_{e \mu} \widetilde{H}_{\mu \tau }
 \widetilde{H}_{\tau e})}{{\rm Im}(H_{e \mu}^M H_{\mu \tau }^M
 H_{\tau e}^M)},
\end{equation}
independently of the baseline of the experiment.  This is one
of the main results of the present work.

{\it On-resonance}: $\widetilde{\Delta}_{31} \simeq 0$.  The near
cancellation of ${\cal O}(\sqrt{2} G_F N_e)$  with other terms in
Eq.\ (\ref{eq:masses}) and the condition (\ref{eq:gammadelta})
together imply for Eq.\ (\ref{eq:masses21}) that
\begin{equation}
 \widetilde{\Delta}_{12} \simeq \frac{m^2_3 - m^2_2}{2E} +
 {\cal O}(2 E |\phi| \delta \gamma) \simeq \Delta_{12},
\end{equation}
and thus
\begin{equation}
\label{eq:delta12}
 \sin \frac{\widetilde{\Delta}_{12} L}{2} \simeq \sin
 \frac{\Delta_{12} L}{2}.
\end{equation}
Furthermore, if the experimental baseline and the neutrino
energy satisfy the condition
\begin{equation}
\label{eq:baselineonres}
 L < \frac{2}{|\widetilde{\Delta}_{31}|} \simeq \frac{2}{{\rm
 Max}\left( \frac{|\delta m^2_{31}|}{2E}\sin 2 \theta_3,  \ 2E
 |\phi \delta \gamma| \right)},
\end{equation}
then the approximations
\begin{equation}
 \label{eq:morestuff} \sin \frac{\widetilde{\Delta}_{31} L}{2}
 \simeq \frac{\widetilde{\Delta}_{31} L}{2}, \qquad \sin
 \frac{\Delta_{31} L}{2} \simeq \frac{\Delta_{31} L}{2}
\end{equation}
are also well-founded.   Substituting Eqs.\ (\ref{eq:delta32}),
(\ref{eq:delta12}) and (\ref{eq:morestuff}) into Eq.\
(\ref{eq:ratio}), we find that the $T$-asymmetry ratio in the
on-resonance case is equally well described by Eq.\
(\ref{eq:ratioapprox}), except for a different baseline
requirement (\ref{eq:baselineonres}).

As it turns out, Eq.\ (\ref{eq:ratioapprox}) is a generally
valid approximation for the preferred parameters of a neutrino
factory: an average neutrino energy of about 10 to 30~GeV
and a baseline roughly between 3000 and 6000~km, provided that
Eq.\ (\ref{eq:gammadelta}) holds.

It is useful to rewrite Eq.\ (\ref{eq:ratioapprox}) as
\begin{equation}
 \frac{\Delta \widetilde{P}_T}{\Delta P_T} \equiv 1 + R,
\end{equation}
with
\begin{equation}
\label{eq:increment}
 R \simeq \sum_{ijk}^{\rm cyclic}
 \left[\frac{{\rm Im}(H^M_{ij} H^M_{jk} H^{G'}_{ki})}{{\rm Im}(H_{e
 \mu}^M H_{\mu \tau }^M H_{\tau e}^M)}+ \frac{{\rm Im}(H^{G'}_{ij}
 H^{G'}_{jk} H^{M}_{ki})}{{\rm Im}(H_{e \mu}^M H_{\mu \tau }^M
 H_{\tau e}^M)}\right],
\end{equation}
where $H^{G'}_{ij} \equiv (W H^G W^{\dagger})_{ij}$, the summation
is taken  over $ijk=e \mu \tau,\ \mu \tau e, \ \tau e \mu$,  and
we have used
\begin{equation}
 {\rm Im}(H^{G'}_{e \mu} H^{G'}_{\mu \tau} H^{G'}_{\tau e}) = {\rm
 Im}(H^{G}_{e \mu} H^{G}_{\mu \tau} H^{G}_{\tau e}) = 0,
\end{equation}
since two of the three gravitational eigenstates are by assumption
degenerate.  The five extra parameters from $H^{G'}$ offer a
plethora of possibilities.  We provide here a few illustrative
cases, but the list is by no means exhaustive.

\begin{figure}
\epsfig{file=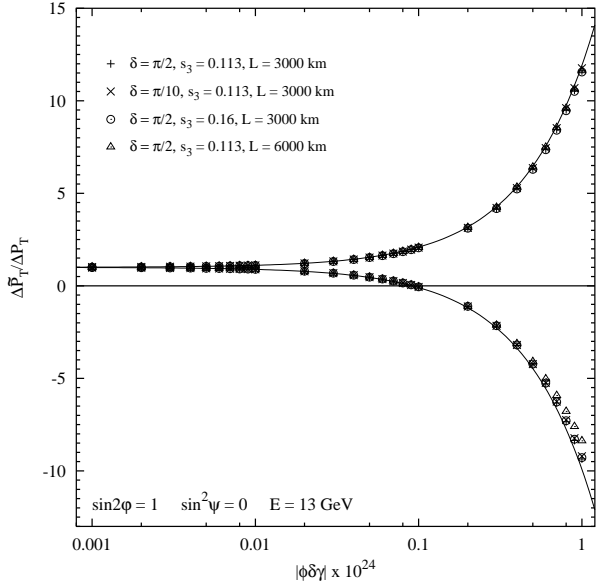, width=8cm} \caption{$\Delta
\widetilde{P}_T/\Delta P_T$ versus $|\phi| \delta \gamma$ for
$E=13\ {\rm GeV}$, maximal $\nu_e \leftrightarrow \nu_{\tau}$ VEP
mixing ($\sin 2 \varphi = 1$ and $\sin^2 \psi=0$), and with the
relative phase $\beta$ set to zero.  The standard oscillation
parameters used here are the best fit values: $|\delta m^2_{32}| =
3 \times 10^{-3}\ {\rm eV}^2$, $\sin 2 \theta_2 = 1$, $\delta
m^2_{21} = 5 \times 10^{-5} {\rm eV}^2$, and $\sin 2 \theta_1=0.8$
. The four sets of points correspond to exact values of $\Delta
\widetilde{P}_T/\Delta P_T$ from numerical calculations for
different values of $s_3$, $\delta$, and baselines $L$.  The solid
lines come from the approximate equation (\ref{eq:etau}); the top
(bottom) set is for negative (positive) $|\phi| \delta
\gamma$'s.\label{fig:etau13gev}}
\end{figure}

\begin{figure}
\epsfig{file=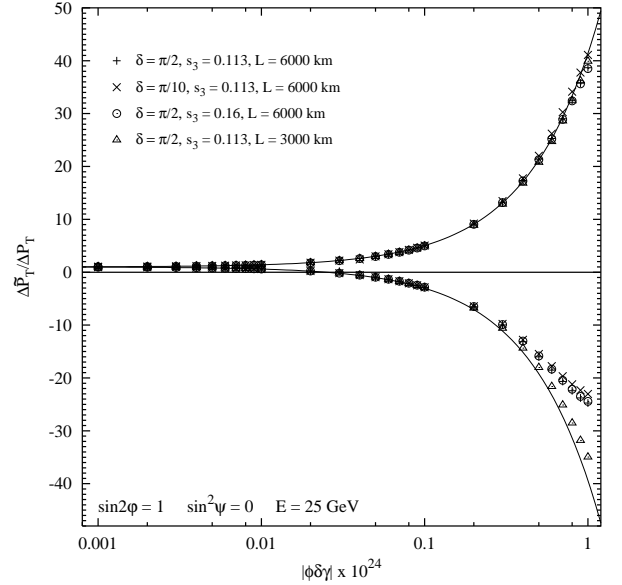, width=8cm} \caption{Same as Fig.\
\ref{fig:etau13gev}, except for $E = 25\ {\rm
GeV}$.\label{fig:etau25gev}}
\end{figure}

\subsection{Pure $\nu_e \leftrightarrow \nu_{\tau}$ VEP mixing}

The setting of $\psi=0$  corresponds to pure $\nu_e
\leftrightarrow \nu_{\tau}$ mixing in the VEP sector.  In Eq.\
(\ref{eq:increment}), the only surviving term is
\begin{equation}
 R \simeq \frac{{\rm Im}(H^M_{e \mu} H^M_{\mu \tau} H^{G'}_{\tau
 e})}{{\rm Im}(H_{e \mu}^M H_{\mu \tau }^M H_{\tau e}^M)}.
\end{equation}
Using the parameterisation introduced in Sec.\
\ref{sec:nomenclature}, we obtain
\begin{equation}
\label{eq:etauratio}
 \frac{\Delta \widetilde{P}_T}{\Delta P_T}  \simeq 1 +
 \frac{2 E^2 B}{\delta m^2_{21}} \ |\phi|\delta \gamma \sin 2
 \varphi,
\end{equation}
where
\begin{equation}
\label{eq:b}
 B   \simeq
 \frac{s_2}{s_1 c_1} \frac{\sin (\beta - \delta)}{\sin
 \delta}\left( 1 + \frac{\delta m^2_{21}}{\delta m^2_{32}} \right)
 +\frac{c_2}{s_3}\frac{\delta m^2_{21}}{\delta m^2_{32}} \frac{\sin
 \beta}{\sin \delta},
\end{equation}
to leading order in $(\delta m^2_{21}/\delta m^2_{32})$ and $s_3$.
As expected, the increment $R$ scales with the square of the
neutrino energy $E^2$ and the VEP breaking scale $|\phi| \delta
\gamma$ which may be positive or negative . Since we have
restricted $|\phi \delta \gamma|$ to be much smaller than the
atmospheric oscillation scale [Eq.\ (\ref{eq:gammadelta})],
$|\phi| \delta \gamma$ interferes substantially only with the
solar neutrino mass splitting $\delta m^2_{21}$, and hence only
$\delta m^2_{21}$ appears in the leading order term of $R$.
Naturally, the degree of $\nu_e \leftrightarrow \nu_{\tau}$ VEP
mixing, $\sin 2 \varphi$, is also a contributing factor.

\begin{figure}
\epsfig{file=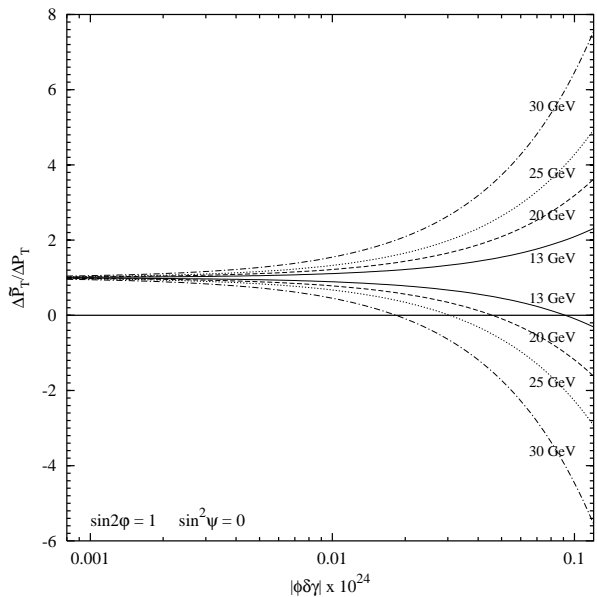,  width=8cm} \caption{$\Delta
\widetilde{P}/\Delta P$ versus $|\phi| \delta \gamma$ for
different neutrino energies.  Maximal $\nu_e \leftrightarrow
\nu_{\tau}$ VEP mixing is assumed, and $\beta=0$. The top (bottom)
set of curves is for negative (positive) $|\phi| \delta \gamma$'s.
We use the best-fit standard oscillation parameters and
$s_3=0.113$. \label{fig:energy}}
\end{figure}

\begin{figure}
\epsfig{file=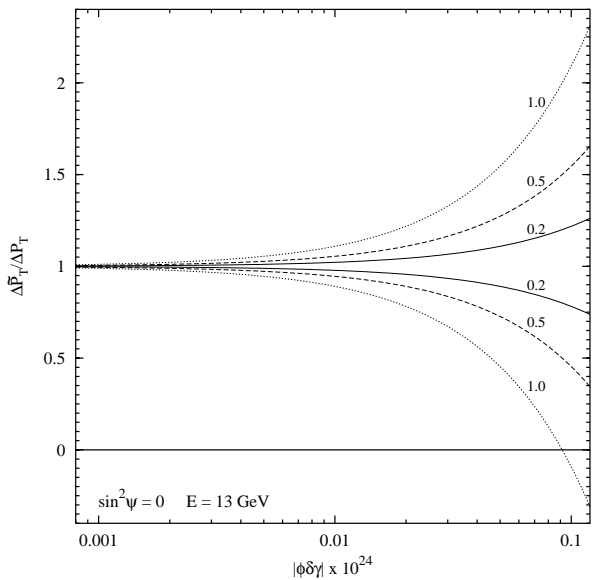, width=8cm} \caption{$\Delta
\widetilde{P}/\Delta P$ versus $|\phi| \delta \gamma$ for
different values of $\sin 2 \varphi$, while $\sin^2 \psi$ and
$\beta$ are set to zero, and the neutrino energy is fixed at $13 \
{\rm GeV}$.  The top (bottom) set of curves is for negative
(positive) $|\phi| \delta \gamma$'s. \label{fig:angles}}
\end{figure}

The quantity $B$ depends on the mixing angles of $U_M$ (some of
which we assume can be established from existing experimental
data), as well as the intrinsic $CP$ phase $\delta$ and one of the
relative phases $\beta$.  In the simplest case where $\beta = 0$,
$B$ becomes independent of both $\theta_3$ and $\delta$ such that
\begin{equation}
\label{eq:etau}
 \frac{\Delta \widetilde{P}_T}{\Delta P_T} \simeq 1
 - \frac{4 E^2}{\delta m^2_{21}} \frac{\sin  \theta_2}{\sin 2
 \theta_1} |\phi| \delta \gamma \sin 2 \varphi,
\end{equation}
to zeroth order in $(\delta m^2_{21}/\delta m^2_{32})$. Figure
\ref{fig:etau13gev} shows the approximate equation (\ref{eq:etau})
plotted as a function of $|\phi| \delta \gamma$ with the neutrino
energy fixed at $13\ {\rm GeV}$ for maximal VEP mixing, $\sin 2
\varphi =1$. Also displayed in juxtaposition are four sets of data
points corresponding to exact values of $\Delta
\widetilde{P}_T/\Delta P_T$ calculated numerically for different
values of $s_3$, $\delta$, and baselines $L$.  Figure
\ref{fig:etau25gev} is similar to Fig.\ \ref{fig:etau13gev}, save
for a different neutrino energy, $E = 25 \ {\rm GeV}$. Concordance
between the exact and the approximate results generally obtains.
Deviations from the exact results appear for the larger $|\phi
\delta \gamma|$'s ($\gtrsim 10^{-25}$) as one tunes up the
neutrino energy, since satisfaction of the requirement
(\ref{eq:gammadelta}) becomes increasingly poor.

\begin{figure}
\epsfig{file=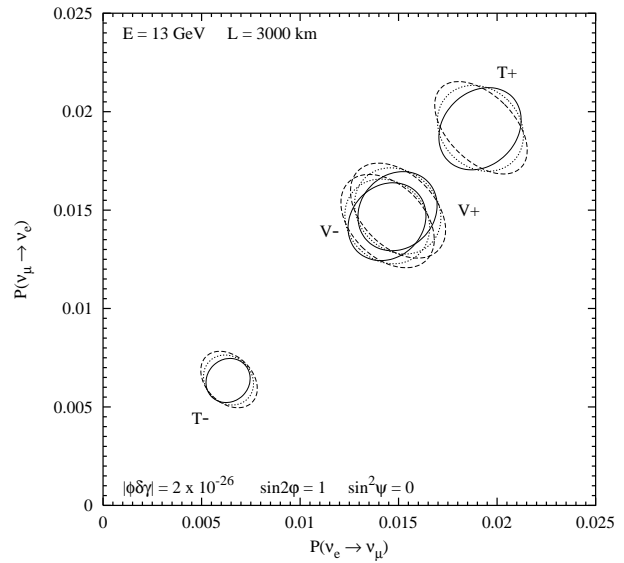, width=8cm} \caption{$T$-trajectory
diagram for maximal $\nu_e \leftrightarrow \nu_{\tau}$ VEP mixing,
with $\beta=0$. The ellipses labelled $V+(-)$ are the trajectories
in vacuum for a positive (negative) $\delta m^2_{32}$. Those
labelled $T+$ and $T-$ are the corresponding trajectories in
matter.  Solid (dashed) lines are for a positive (negative)
$|\phi| \delta \gamma$, while dotted lines are for the standard,
no VEP case. \label{fig:etautraj}}
\end{figure}

Depending on the relative sign of $|\phi| \delta \gamma$ and
$\delta m^2_{21}$, the ratio $\Delta \widetilde{P}_T/\Delta P_T$
splits into two branches for any given set of $E$ and $\sin 2
\varphi$. Since the sign of $\delta m^2_{21}$ is fixed by the
solar neutrino data to be positive, the dependence is on the sign
of $|\phi| \delta \gamma$ only, as is evident in Figs.\
\ref{fig:etau13gev} and \ref{fig:etau25gev}. The top branch, where
$|\phi| \delta \gamma <0$, always leads to an enhancement in the
observed $T$-asymmetry. The bottom branch, on the other hand,
shows a $\Delta \widetilde{P}_T/\Delta P_T$ that tends to zero,
flips sign, and then continues to grow in magnitude as we increase
$|\phi \delta \gamma|$.

\begin{figure}
\epsfig{file=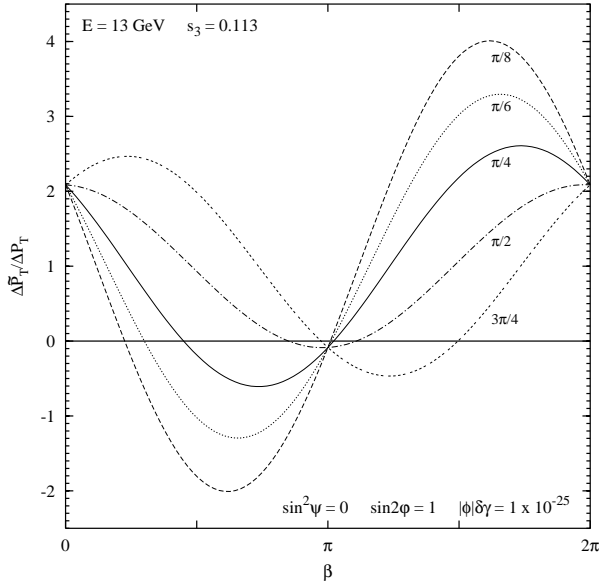,  width=8cm} \caption{$\Delta
\widetilde{P}_T/\Delta P_T$ versus $\beta$ for different values of
$\delta$ for maximal $\nu_e \leftrightarrow \nu_{\tau}$ VEP
mixing.  The neutrino energy is fixed at $13 \ {\rm GeV}$ and
$|\phi| \delta \gamma = 1 \times 10^{-25}$. \label{fig:phase}}
\end{figure}

More $\Delta \widetilde{P}_T/\Delta P_T$ versus $|\phi| \delta
\gamma$ curves are displayed in Figs.\ \ref{fig:energy} and
\ref{fig:angles}, for different neutrino energies and mixing
parameters $\sin 2 \varphi$.  Figure \ref{fig:energy} shows that
deviations of $\Delta \widetilde{P}_T$ from $\Delta P_T$ are more
prominent for higher energy neutrinos, which is consistent with
expectation. Depending on the neutrino energy, the enhancement in
the magnitude of the $T$-asymmetry in the parameter region
$10^{-26} \lesssim |\phi \delta \gamma| \lesssim 10^{-25}$ can be
anywhere from $\sim 20\ \%$ to a factor of a few.

\begin{figure}
\epsfig{file=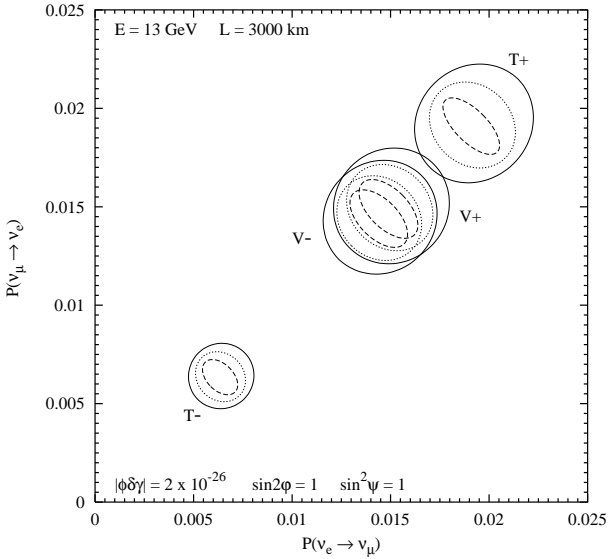,  width=8cm} \caption{Same as Fig.\
\ref{fig:etautraj}, but for maximal $\nu_e \leftrightarrow
\nu_{\mu}$ VEP mixing and $\alpha =0$.\label{fig:emutraj}}
\end{figure}

It is interesting to note that the corresponding changes in the
actual oscillation probabilities are generally no more than a few
percent, as illustrated, for example, by the $T$-trajectory
diagrams in Fig.\ \ref{fig:etautraj}.\footnote{$CP$- and
$T$-trajectory diagrams were first introduced in Refs.\
\cite{bib:minakata} and \cite{bib:parke} respectively.} Thus, by
comparing the appearance rates of two $T$-conjugate channels, we
will be able to probe a whole new region of VEP parameter space
that is inaccessible to conventional single channel appearance or
disappearance experiments run at the same energies.

\begin{figure}
\epsfig{file=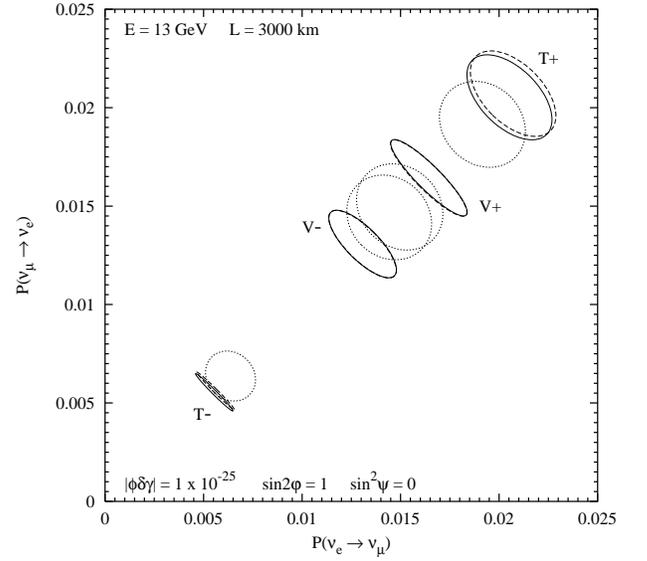,  width=8cm}\caption{$T$-trajectory diagram
for maximal $\nu_e \leftrightarrow \nu_{\tau}$ VEP mixing with
$\beta$ as the phase variable.  The standard $CP$ phase $\delta$
is set to identically zero. The ellipses labelled $V+(-)$ are the
trajectories in vacuum for a positive (negative) $\delta
m^2_{32}$. Those labelled $T+$ and $T-$ are the corresponding
trajectories in matter.  Solid (dashed) lines are for a positive
(negative) $|\phi| \delta \gamma$. For the purpose of comparison,
we have also included the standard $T$-trajectories (as functions
of $\delta$).  These are represented by the dotted
lines.\label{fig:betatraj}}
\end{figure}

Before concluding this section, let us remark on the relative
phase $\beta$. Figure \ref{fig:phase} shows $\Delta
\widetilde{P}_T/\Delta P_T$ as a function of $\beta$, according to
Eqs.\ (\ref{eq:etauratio}) and (\ref{eq:b}), for several values of
$\delta$'s at a fixed $E^2 |\phi| \delta \gamma \sin 2 \varphi$.
The message here is clear: extra phases in the Hamiltonian do not
always contribute to increasing the extent of $T$-violation,
particularly when $\delta$ is already close to the maximally
$T$-violating value of $\pi/2$ or $3 \pi/2$. Interference between
the phases means that the final $T$-asymmetry may even be
suppressed to zero. On the other hand, if $|\sin \delta| \ll |\sin
\beta|$, the latter becomes the dominant phase, and the
$T$-asymmetry generated therefrom peaks at $\beta \simeq \pi/2, 3
\pi/2$, as is evident in Fig.\ \ref{fig:phase}. We will revisit
this issue in a later section.

\subsection{Pure $\nu_e \leftrightarrow \nu_{\mu}$ VEP mixing}

This case corresponds to setting $\psi = \pi/2$.  The
$T$-asymmetry ratio is given by
\begin{equation}
\label{eq:emuratio}
 \frac{\Delta \widetilde{P}_T}{\Delta P_T}
 \simeq 1 + \frac{2 E^2 A}{\delta m^2_{21}}\  |\phi|\delta \gamma
 \sin 2 \varphi,
\end{equation}
where
\begin{equation}
\label{eq:a}
 A  \simeq  \frac{c_2}{s_1 c_1} \frac{\sin (\delta -
 \alpha)}{\sin \delta} \left( 1 + \frac{\delta m^2_{21}}{\delta
 m^2_{32}} \right) + \frac{s_2}{s_3}\frac{\delta m^2_{21}}{\delta
 m^2_{32}} \frac{\sin \alpha}{\sin \delta},
\end{equation}
to leading order in $(\delta m^2_{21}/\delta m^2_{32})$ and $s_3$.
Comparing Eqs.\ (\ref{eq:b}) and (\ref{eq:a}), we see that since
$s_2 \simeq c_2 \simeq 1/\sqrt{2}$, the effects of this VEP mixing
channel on the $T$-asymmetry must be very similar to those from
pure $\nu_e \leftrightarrow \nu_{\tau}$ VEP mixing.  In the case
of $\alpha=\beta=0$, the changes induced in $\Delta
\widetilde{P}_T$ are identical for the two mixing channels upon
replacing $|\phi| \delta \gamma \to - |\phi| \delta \gamma$ in
Eq.\ (\ref{eq:emuratio}). The actual oscillation probabilities,
however, are quite different. Compare the $T$-trajectories in
Fig.\ \ref{fig:emutraj} with those in Fig.\ \ref{fig:etautraj}.

\subsection{Pure $\nu_{\mu} \leftrightarrow \nu_{\tau}$ VEP mixing}

The case of $\nu_{\mu} \leftrightarrow \nu_{\tau}$ mixing is
equivalent to setting $\varphi=0$, such that
\begin{equation}
 \frac{\Delta \widetilde{P}_T}{\Delta P_T}  \simeq 1 +
 \frac{2 E^2 C}{\delta m^2_{21}} \ |\phi|\delta \gamma \sin 2 \psi,
\end{equation}
where
\begin{eqnarray}
\label{eq:c}
 C \!\!& \simeq & \! \! - \frac{1}{s_2
 c_2}\frac{\delta m^2_{21}}{\delta m^2_{32}} \frac{s_2^2
 \sin(\alpha - \beta - \delta) - c_2^2 \sin(\alpha - \beta +
 \delta)}{\sin \delta} \nonumber \\ && \hspace{5mm} +
 \frac{s_3}{s_1 c_1} \frac{\sin(\alpha - \beta)}{\sin \delta}
\end{eqnarray}
is suppressed by a factor of $(\delta m^2_{21}/\delta m^2_{32})$
or $s_3$.  This implies that the magnitude of the increment $R$ is
generally some ${\cal O} (10^{-1}) \to {\cal O} (10^{-2})$ times
smaller than would be observed in the case of, say, pure $\nu_e
\leftrightarrow \nu_{\tau}$ VEP mixing for the same $|\phi| \delta
\gamma$ and mixing parameter, and at the same energy.  At $E = 13
\ {\rm GeV}$ and $|\phi \delta \gamma| \sim 10^{-25}$, the change
in the $T$-asymmetry is no more than one per cent, unless $|\sin
\delta| \ll |\sin (\alpha - \beta)|$. Thus $T$-violation
experiments run at energies of a few tens GeV are unable to probe
VEP in the $(\nu_{\mu},\nu_{\tau})$ sector to below $\sim
10^{-24}$, let alone supersede the $|\phi \delta \gamma| \lesssim
10^{-26}$ bound imposed by the Super-Kamiokande atmospheric
neutrino data \cite{bib:italian}.

\section{$T$-violation with $\delta = 0$}

An interesting consequence of a nonzero $\alpha$ and/or $\beta$ is
that either or both of these relative phases alone can give rise
to $T$-violation in the oscillation probabilities, in the absence
of any explicit $CP$- or $T$-violating phases in the mixing
matrices $U_M'$ and $U_G'$.  Consider for example VEP mixing only
in the $(\nu_e, \nu_{\tau})$ sector, i.e., $\psi=0$. From Eqs.\
(\ref{eq:etauratio}) and (\ref{eq:b}), the asymmetry $\Delta
\widetilde{P}_T$ is given approximately by
\begin{eqnarray}
 \Delta \widetilde{P}_T \!\!  &\simeq&\! \! Q \sin \beta \!\times\!
 \left\{\! 16  s_1 c_1 s_2 c_2 s_3
 c_3^2 \left[ \frac{1}{8 E^3} \frac{\delta m^2_{12} \delta m^2_{23}
 \delta m^2_{31}}{\Delta_{12} \Delta_{23} \Delta_{31}} \right]
 \right. \nonumber
 \\ && \hspace{12mm} \left.  \times \sin \! \frac{\Delta_{12} L}
 {2}\ \sin \!\frac{\Delta_{23} L}{2}\ \sin \! \frac{\Delta_{31} L}
 {2} \right\},
\end{eqnarray}
where
\begin{equation}
 Q = \frac{2 E^2 |\phi| \delta \gamma \sin 2 \varphi}{\delta
 m^2_{21}} \!\left[\!\frac{s_2}{s_1 c_1}\! \left( 1 \!+\!
 \frac{\delta m^2_{21}}{\delta m^2_{32}}  \right) \!+\!
 \frac{c_2}{s_3} \frac{\delta m^2_{21}}{\delta m^2_{32}} \right],
\end{equation}
in the limit $\delta \to 0$. Comparing this with the standard
$T$-asymmetry,
\begin{eqnarray}
 \Delta P_T \!\!  &=&\! \! \sin \delta \times \left\{16 \ s_1 c_1
 s_2 c_2 s_3 c_3^2 \left[ \frac{1}{8 E^3} \frac{\delta m^2_{12}
 \delta m^2_{23} \delta m^2_{31}}{\Delta_{12} \Delta_{23}
 \Delta_{31}} \right] \right. \nonumber
 \\ && \hspace{12mm} \left.  \times \sin \! \frac{\Delta_{12} L}
 {2}\ \sin \!\frac{\Delta_{23} L}{2}\ \sin \! \frac{\Delta_{31} L}
 {2} \right\},
\end{eqnarray}
we see that the relative phase $\beta$ can mimic the role of the
standard phase $\delta$.

Assuming $\delta=0$, Fig.\ \ref{fig:betatraj} shows the
$T$-trajectories  for this case at $E = 13\ {\rm GeV}$ and $|\phi
\delta \gamma| = 1 \times 10^{-25}$, with $\beta$ as the phase
variable.  Because of the factor $Q$, the shape of these
$T$-trajectories can be quite different from the standard
$T$-trajectories (as functions of $\delta$), which are displayed
also in the same figure.

\begin{figure}
\epsfig{file=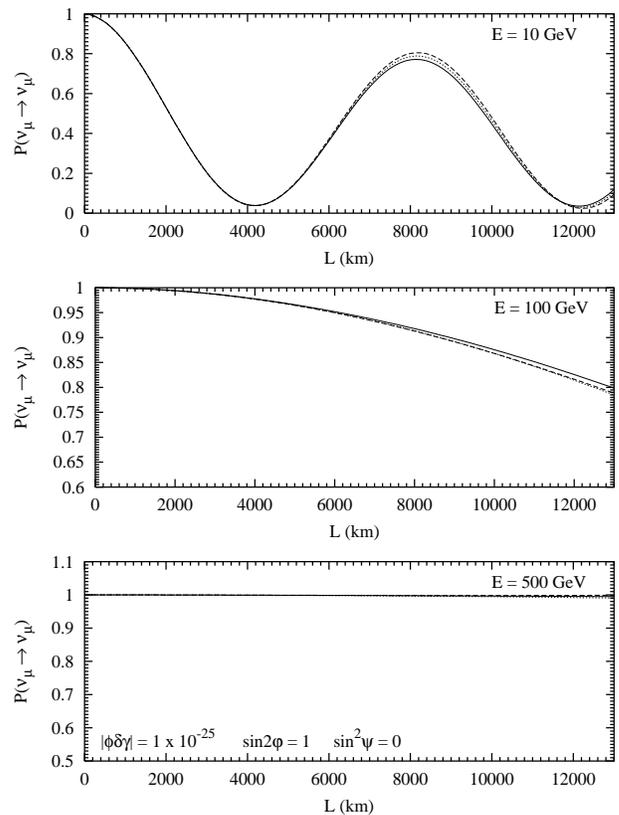, width=8cm} \caption{$\nu_{\mu}$ survival
probability at various energies and $|\phi \delta \gamma| = 1
\times 10^{-25}$. Solid (dashed) lines are for maximal $\nu_e
\leftrightarrow \nu_{\tau}$ VEP mixing with a positive (negative)
$|\phi| \delta \gamma$. Dotted lines are for the standard, no VEP
case.\label{fig:etausk}}
\end{figure}

\section{Compatibility with the Super-Kamiokande atmospheric neutrino data}

Since the sensitivity of an experiment to VEP-LIV increases with
$L \cdot E$, it is instructive to reexamine our case in the light of
the Super-Kamiokande atmospheric neutrino data, whose parent neutrino
distribution spans some four decades in neutrino energy, $E \sim
0.1 \to 10^{3} \ {\rm GeV}$.  We do not have the technology to
perform a full-fledged three-flavour fit to the data to ensure
that the VEP parameters considered in this work are not in
conflict with experimental data.  However, the same purpose may be
more than amply achieved by requiring the new $\nu_{\mu}$ survival
probabilities not to deviate too much from that given by the best
fit parameters.

\begin{figure}
\epsfig{file=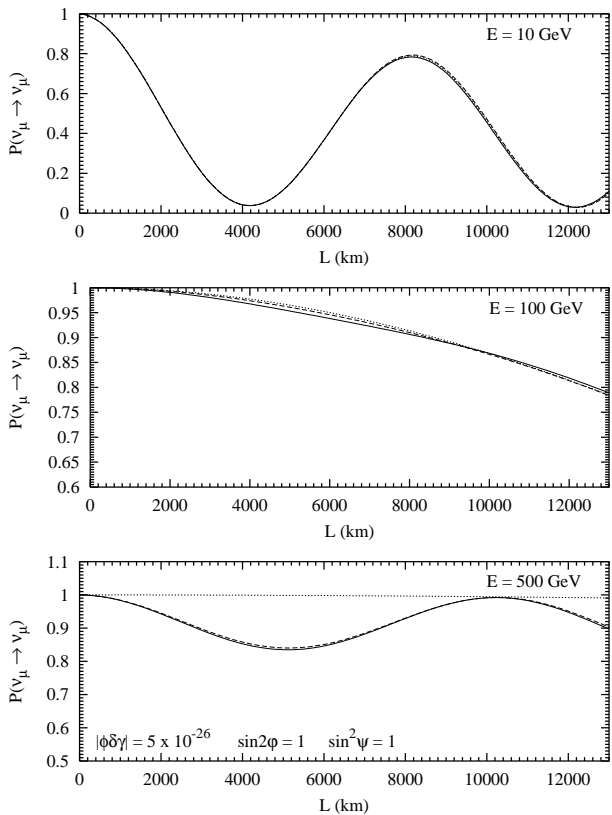, width=8cm} \caption{Same as Fig.\
\ref{fig:etausk}, but for $|\phi \delta \gamma| = 5 \times
10^{-26}$ and maximal $\nu_{e} \leftrightarrow \nu_{\mu}$ VEP
mixing.\label{fig:emusk}}
\end{figure}

\begin{figure}
\epsfig{file=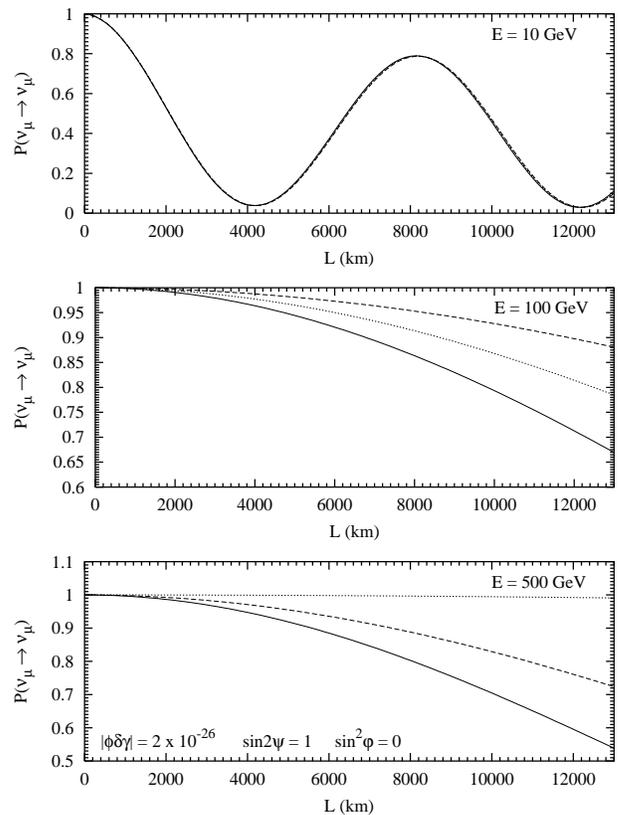, width=8cm} \caption{Same as Fig.\
\ref{fig:etausk}, but for $|\phi \delta \gamma| = 2 \times
10^{-26}$ and maximal $\nu_{\mu} \leftrightarrow \nu_{\tau}$ VEP
mixing.  According to Ref.\ \cite{bib:italian}, the solid and
dashed curves are allowed by the Super-Kamiokande atmospheric
neutrino data.\label{fig:mutausk}}
\end{figure}

Figures \ref{fig:etausk} and \ref{fig:emusk} show the $\nu_{\mu}$
survival probability at several energies, with maximal VEP mixing
in the $(\nu_e, \nu_{\tau})$ and $(\nu_{e}, \nu_{\mu})$ sectors
respectively.  In the case of $\nu_e \leftrightarrow \nu_{\tau}$
VEP mixing, there is virtually no deviation from the standard
$\nu_{\mu}$ survival probability at $E= 10,\ 100$, and $500\ {\rm
GeV}$ for $|\phi \delta \gamma| = 1 \times 10^{-25}$.  The case of
$\nu_{e} \leftrightarrow \nu_{\mu}$ VEP mixing seems somewhat more
restricted.  At $E=500\ {\rm GeV}$ and $|\phi \delta \gamma| = 5
\times 10^{-26}$, the departure from the standard probability can
be as large as $\sim 15 \ \%$.  However, bearing in mind that the
neutrino source peaks at $\sim 100\ {\rm GeV}$, this divergence at
$500\ {\rm GeV}$ is most likely still acceptable.  The case of VEP
mixing in the $(\nu_{\mu}, \nu_{\tau})$ sector was considered in
Ref.\ \cite{bib:italian} in a two-flavour analysis. For the
purpose of comparison, we plot in Fig.\ \ref{fig:mutausk} the
$\nu_{\mu}$ survival probability for a $|\phi \delta \gamma|$ just
above the upper limit of $10^{-26}$ from the said analysis.
Clearly, the curves in Figs.\ \ref{fig:etausk} and \ref{fig:emusk}
are no worse than these ``acceptable'' ones.  Thus we can be quite
confident that the VEP parameters considered in this work have not
already been ruled out by Super-Kamiokande.  A proper analysis is
required to settle this issue.

\section{Conclusion}

We have considered in this work the prospects for extending the
limits on equivalence principle and Lorentz invariance violations
(VEP and LIV) in future neutrino factory experiments.  We have
examined three neutrino flavour oscillations, treating VEP-LIV as
a sub-leading mechanism to the standard mass-mixing mechanism,
under the simplifying assumption that the neutrino-gravity
couplings of two of the three neutrino gravitational eigenstates
are nearly the same. The presence of nondegenerate
neutrino-gravity couplings introduces additional mixing and phases
which can lead to $CP$- and $T$-violation even if the phase
$\delta$ in the MNS mixing matrix vanishes. $T$-violation
measurements have the potential to yield the most stringent bounds
on VEP-LIV.  Our case study shows that, with suitable neutrino
path lengths of several thousand kilometres, the limits on VEP-LIV
for the $(\nu_e, \nu_\mu)$ and $(\nu_e, \nu_\tau)$ sectors can be
lowered to $|\phi \delta \gamma| \lesssim 10^{-26}$, which matches
the limit for the $(\nu_\mu, \nu_\tau)$ sector imposed by the
Super-Kamiokande atmospheric neutrino data. At present, we see
little prospect for extending this limit much beyond the
$10^{-26}$ level.

\acknowledgments{This work was supported in part by the U. S.
Department of Energy under grant DE-FG02-84ER40163.  C.N.L.\ would
like to thank S.~M.~Barr, I.~Dorsner, and A.~Halprin for useful
discussions.}

\end{document}